\documentclass[aps, twocolumn, tightenlines,amscd,amsmath,amssymb,verbatim,nofootinbib]{revtex4-1}

\usepackage{graphicx}
\usepackage{subfig}
\begin{document}

\title{Modeling the adoption of innovations in the presence of geographic and media influences}
\author{Jameson L. Toole*}
\affiliation{Engineering Systems Division, Massachusetts Institute of Technology, Cambridge, MA}
\author{Meeyoung Cha}
\affiliation{Graduate School of Culture Technology, Korea Advanced Institute of Science and Technology, Korea}
\author{Marta C. Gonz\'{a}lez}
\affiliation{Department of Civil and Environmental Engineering, Massachusetts Institute of Technology, Cambridge MA}

\begin{abstract}
While there has been much work examining the affects of social network structure on innovation adoption, models to date have lacked important features such as meta-populations reflecting real geography or influence from mass media forces.  In this article, we show these are features crucial to producing more accurate predictions of a social contagion and technology adoption at the city level.  Using data from the adoption of the popular micro-blogging platform, Twitter, we present a model of adoption on a network that places friendships in real geographic space and exposes individuals to mass media influence.  We show that homopholy both amongst individuals with similar propensities to adopt a technology and geographic location are critical to reproduce features of real spatiotemporal adoption.  Furthermore, we estimate that mass media was responsible for increasing Twitter's user base two to four fold.  To reflect this strength, we extend traditional contagion models to include an endogenous mass media agent that responds to those adopting an innovation as well as influencing agents to adopt themselves.
\end{abstract}

\maketitle
In an increasingly digital and connected world, the processes by which information is shared and consumed are changing rapidly.  Services and content are now distributed through on-line social networks where the flattening affects of the Internet distort spatial diffusion.  These factors are quickly shifting the balance between word-of-mouth contagion and more traditional mass media advertisement, changing the spatio-temporal scales on which spreading occurs.  Aiding our ability to characterize and quantify this shift are unprecedented amounts of data elucidating how people communicate with each other and how that communication translates in choices or behaviors such as adopting an innovation or technology.

In this article, we update and unify traditional models of information spread and technology adoption to more accurately reflect the novel economic and social environments in which the spreading occurs.  We expand on metapopulation models by embedding social networks in real geography to reflect the spatial distribution of social ties and better understand how local demographics and topology affect contagion.  Furthermore, we introduce an endogenous media agent to a network model of information spread, capturing the role of hyper-influential social forces.  Our model is informed by a case study examining the {\em viral} (as it is colloquially referred) adoption of a social micro-blogging platform, Twitter, where we examine the accumulation of users in cities across the country over a period of three years.

Traditional models of contagion have generally focused on the spread of disease \cite{Dietz1967} or the diffusion of innovation \cite{Rogers1995, Valente1995}.  Simple approaches such as the susceptible - infected (SI) model, involving differential equations have proved extremely informative, but suffer from simple assumptions such as homogeneous mixing of a population.  The diffusion of innovations literature has had made use of similar frameworks, such as the Bass model \cite{Bass1969}, to characterize the adoption of technologies with considerable cost and risk.  We show, however, that these models perform poorly when applied to goods and services that are free and demonstrate massive positive externalities such as social web applications.

These processes have been placed on networks, revealing how the topology of our social connections aids or hinders outbreaks.  Findings of this work are especially important when considering a world that becomes increasingly connected by the availability of Internet connections or cheap and fast travel by cars, trains, and planes \cite{Dodds2005, Newman2002, Newman2010, Sterman2008, Fortunato2010}.  Few, however, have placed such networks in real geography while preserving individual interactions, thinking carefully about properties such as homopholy \cite{Watts2005} \cite{Uzzi2008}.

Massive popular interest in social networks has recently lead scholars to recognize the potential of using social network websites for research, where most of these studies until today have focused on spreading of information. For example, it has been shown that different types of information, be it political or related to sports, follow different patterns as they are shared and consumed by millions of individuals \cite{Kleinberg2008, Klienberg2010}.  Some information even takes on a life of its own, evolving into self-sustaining `memes' \cite{Kleinberg2009}.  In many cases, however, predicting the outcomes of such processes has proven extremely difficult \cite{Salganik2006}.

Social scientists have used similar frameworks to study collective action in the form of binary decisions in order to understand a wide variety of phenomena.  Neighborhood segregation \cite{Shelling1973}, riots, technology adoption \cite{Granovetter1973}, and standards setting are just a few examples of behavioral contagion studied \cite{Centola2007, Centola2005, Centola2010, Watts2007, Watts2008}.

More recently, studies have explored the many forces influencing the speed and success of information spreading such as blogs and traditional news outlets \cite{Leskovec2011, Leskovec2007, Klienberg2010}.  These studies have revealed a number of patterns whereby mass media drives conversation on social networks or vice versa.  Finally, marketers and retailers have examined the various roles of celebrity endorsements as well as spatial diffusion of information about products and services in an attempt to optimize business outcomes \cite{Pease2008, Baum2006, Fowler2008, Bass1969, Garber2004, Allaway2003, Onnela2010}.

In this article, we address significant gaps in the above literature.  Namely, we show how the geographic distribution of individuals' differing propensities to adopt (such as early versus late adopters), combined with a preference for friendship with others who share similar tastes and geographic locations, are crucial features to accurately describe micro (at the city level) and macro (at the national level) adoption trends.  Furthermore, we propose a model that includes an endogenous mass media agent that responds to adoption patterns of users as well as influences individuals to adopt an innovation. Based on adoption data from the popular social blogging platform, Twitter, we present our model of contagion to capture salient features. The remainder of this article is organized into three parts: ($i$) we present analysis of the spatiotemporal adoption of Twitter as a case study, examining the roles of word-of-mouth spreading as well as mass media, ($ii$) we use insights from the case study to construct a network model and simulate adoption, ($iii$) and finally we present and discuss results and important parameters of our model.

\section{A Case Study of Twitter}
As of December, 2010, the social micro-blogging platform, Twitter, had amassed roughly 190 million users globally, nearly 80 million of which were located in the United States.  Started in San Francisco in early March, 2006, Twitter epitomizes the speed and efficiency with which an innovation is adopted by a population as well as its power to transform how we communicate.  To achieve its massive success, Twitter relied almost entirely on word-of-mouth spreading during the first two years of its existence, after which buzz began to circulate from traditional news and media outlets.

To understand the adoption of Twitter in both space and time as well as the role of media, we analyze data from the first 3.5 million users to sign up for Twitter in the United States.  From just weeks after its launch (late March, 2006) through its first massive surge in popularity (August, 2009) the data contains both the time and geocoded, self-reported location, at the city level, at which each user account was created.  In total, users signed up in roughly 16,000 unique cities across the country.  We restrict ourselves, however, to cities where at least 1000 users had signed up over the 3 years to ensure that time series are populated.  This left 408 cities to work with.  These 408 cities account for ~2.3 of the ~3.5 million, or roughly 65\% of all users.  For the remainder of this analysis, we restrict ourselves to this thresholded data set. More information on this data set is available in Cha et al. \cite{Cha2010}.

\subsection{Descriptive Statistics}
\begin{figure} 
\centering
	\includegraphics[width=1\linewidth]{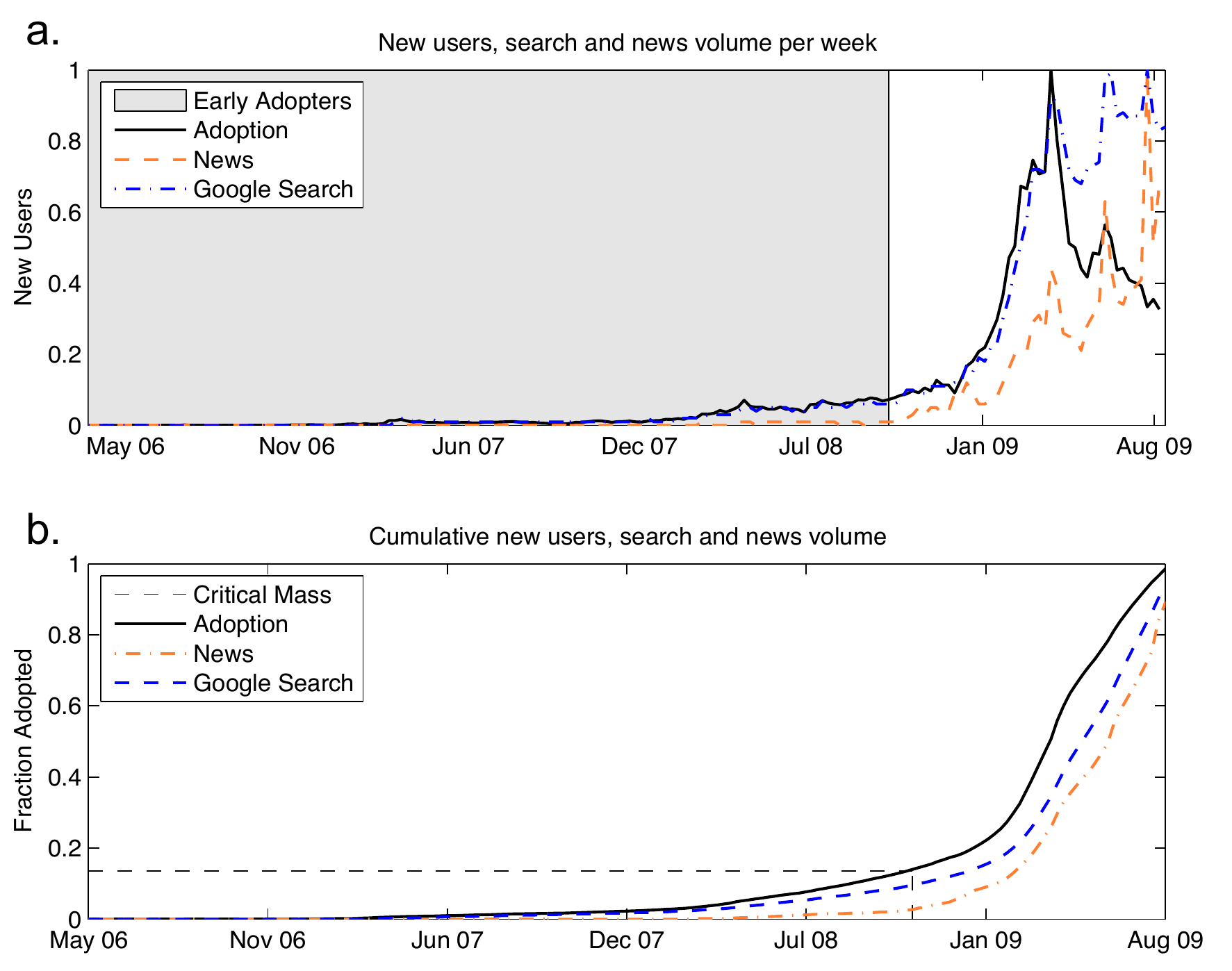} 
	\caption{(a.) The number of new U.S. Twitter users is plotted for each week, normalized by the maximum over the people from mid-March 2006 through late-August 2009.  (b.) The cumulative total number of U.S. Twitter users is plotted for for the same time period.  Google search and news volumes are normalized so that the maximum value is 1.}\label{fig:AggWeekTS}
\end{figure}

As with most complex systems, there are many different scales at which to analyze dynamics.  We start by counting the number of new users that signed up for Twitter across the entire country each week and plot both the week-to-week increase as well as the cumulative sum of all users. In addition, we have gathered data from Google's Trends and Insights web application measuring weekly search volume  news reference volume for the query ``Twitter" on Google News\footnotetext{Information on how search and news values are scaled can be found at http://www.google.com/intl/en/trends/about.html} (Fig. \ref{fig:AggWeekTS}).

\begin{figure} 
\centering
	\includegraphics[width=1\linewidth]{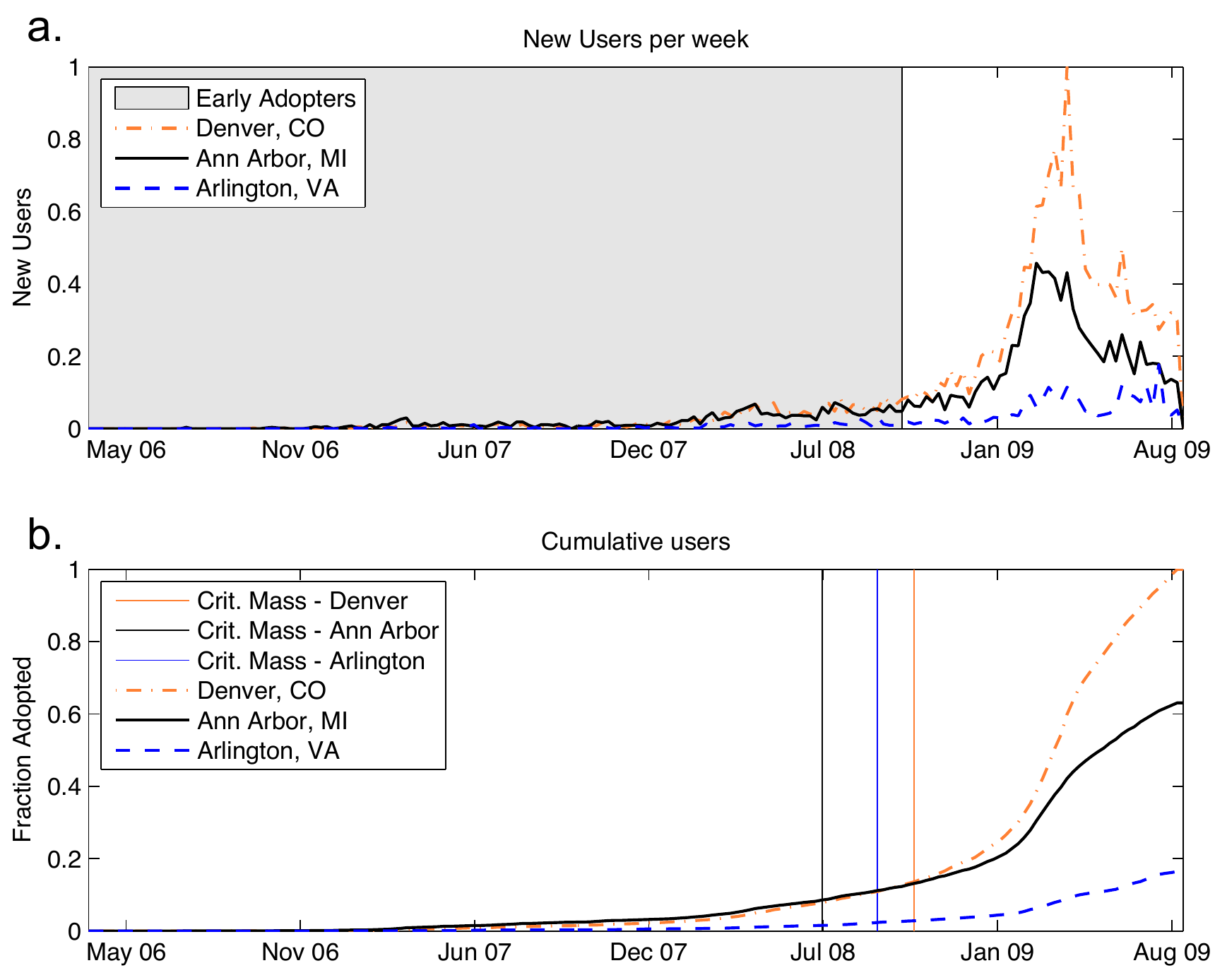} 
	\caption{(a.) Time series display the number of new U.S. Twitter users for three separate locations (Ann Arbor, MI,  Denver, CO, and Arlington, VA) from mid-March 2006 through late-August 2009, normalized by the largest increase in Denver users.  (b.) Shows a plot of the cumulative fraction of each cities user base normalized by Denver, CO total users.}\label{fig:SelectWeekTS}
\end{figure}

Following diffusion of innovations literature we label adopters according to where their adoption times fall relative to the distribution of all other adoption times.  Those who adopt greater than $1\sigma$ (standard deviation)  before the average adoption time are labeled as early adopters.  Those adopting between $1\sigma$ before and the mean adoption time are the early majority, with the late majority and laggards adopting in similar intervals after the mean time.  For more on the motivations behind this, see Rogers \cite{Rogers1995}. Fig. \ref{fig:SelectWeekTS} shows three separate locations across the country representing a young, early adopting demographic (Ann Arbor, MI), a large metropolitan consisting mostly of late majority adopters (Denver, CO), and a mixed area (Arlington, VA).

We measure the composition of each city in terms of the percentage of users who are early adopters, early majority, late majority, or laggards.  This type of analysis also serves to normalize locations with respect to population.  We find that cities with the most early adopters tend to have large universities or are technology centers that tend to attract large numbers of young, tech-savvy persons who are likely to adopt  social web applications.  Later, we show that the empirical composition of cities and the demographics they represent is critical to reproducing spatiotemporal diffusion patterns.

\begin{figure*} 
\centering
	\hspace*{.5in}
	\includegraphics[width=.9\linewidth]{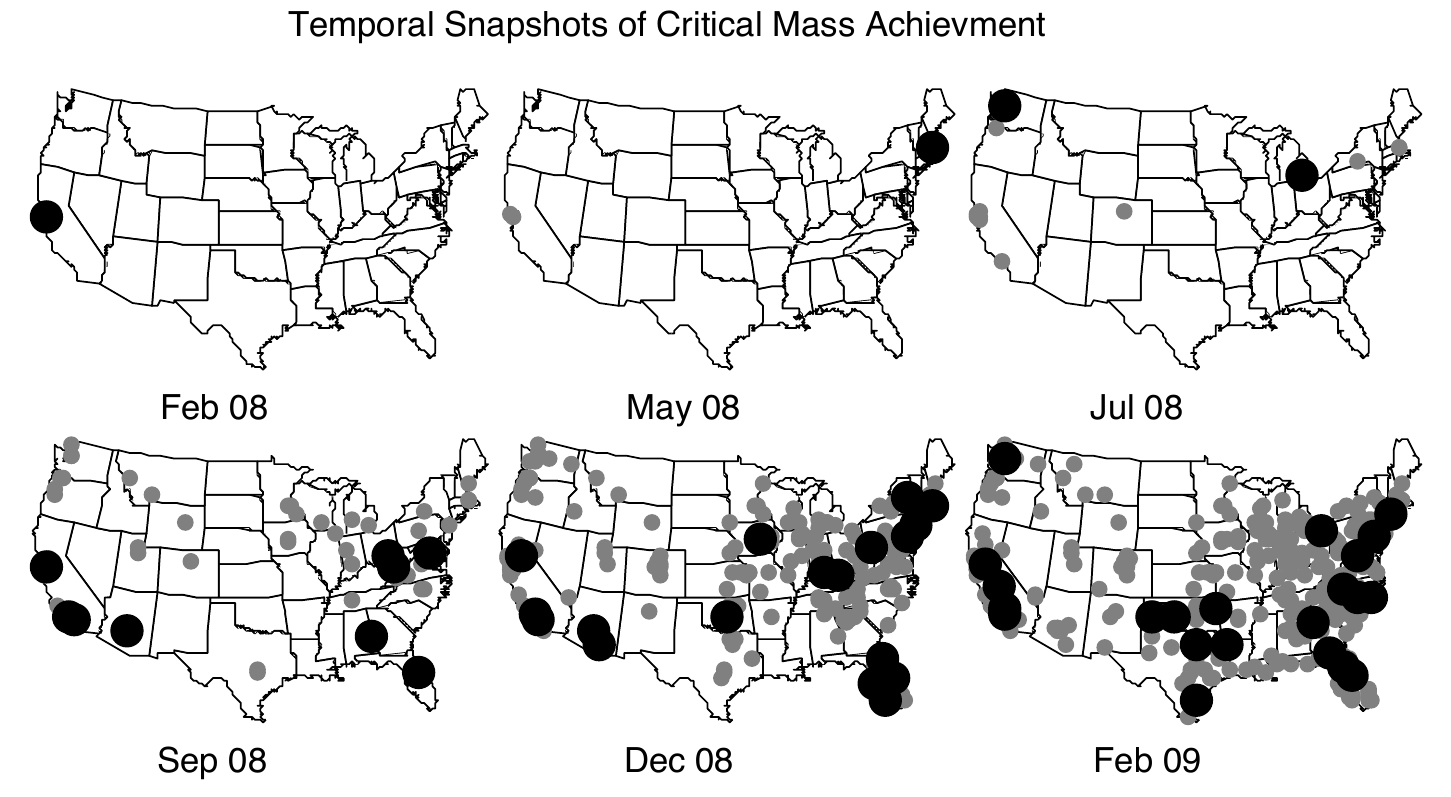} 
	\caption{Temporal snapshots of critical mass achievement at locations across the US.  For each time, the smaller, gray markers indicate locations that have already reached critical mass at that time.  The larger, black markers denote locations that achieved critical mass during that week.  We note that locations achieving critical mass at very early times are clustered around Twitter's birthplace, San Francisco, CA, suggesting local word-of-mouth diffusion.  There are, however, a few locations on the other side of the country, namely the suburbs of Boston, MA that are equally early in adoption, contrasting local diffusion with the flattening affects of the Internet.}\label{fig:CritMassSnapshots}
\end{figure*}

We next focus on a key moment for any contagion process, the critical mass achievement.  Again following the diffusion of innovations literature, we mark a city reaching critical mass when $13.5\%$ of all eventual users have signed up \cite{Valente1995}.  Fig. \ref{fig:CritMassSnapshots} shows a series of snapshots in time indicating when various cities reach critical mass.  These snapshots reveal the diffusion path of Twitter from its birthplace in Silicon Valley, to college towns such as Cambridge, MA, Ann Arbor, MI, or Austin, TX, to metropolitan areas such as Los Angeles, CA, or Denver, CO, then finally to more suburban and rural areas.

Just as individuals users were labeled as an early adopter or a laggard, cities were also placed into groups according to when they reached critical mass relative to the entire population.  Table 1 in the on-line supplementary information displays a complete list cities and their classification to illustrate, qualitatively, the type of demographic information that can be inferred from looking at the adoption of web applications.

Finally, note that media coverage (Fig. \ref{fig:AggWeekTS}) of Twitter was nearly non-existent during the first two years.  During this time, Google search volume was highly correlated with user growth.  After a critical mass of users was reached, the media coverage began to increase super-linearly.  Many of the spikes in adoption rates were the result of celebrity endorsements (Oprah official signed up for Twitter line on air April 17th, 2009) and political events (the Iranian protests in July and August 2009).

Qualitatively, we recognize the media as having an enormous role in driving adoption.  We also find that news coverage did not pick up until after the nation had achieved a critical mass of users, suggesting strong endogeneity where media responds to the very adoption it produces.  This is much different than the traditional modeling of media \cite{Bass1969, Katz1957}.  We seek to capture these stylized facts by including a powerful media agent whose coverage both grows with adoption and produces powerful and random shocks, simulating hyper-influentials and major media events.

\section{ Model and Simulation }
We now introduce our model as follows:

($i$) Contagion spreading is simulated by a mechanism resembling the susceptible - infected (SI) model, which is also a special case of the Bass model, widely used in the diffusion of innovations literature.  We create a population of $N$ agents and place each agent into one of $L$ cities, creating city level meta-populations.  Each agent can be one of two types, {\em early adopter} or {\em regular adopter}.  The geographic placement is chosen to reflect empirical distributions of real Twitter users as well the composition of agent types in each city as discussed in previous sections.  Agents are then connected by links to form a social network.  The empirical characteristics of links and distances can be set to reflect those measured in on-line social networks.  For example, Liben-Nowell et al. ~\cite{Nowell2005} shwo that $p_r$, the probability of being connected to someone located a distance $r$ from your city, following a truncated power-law, $p_r = r^{-\gamma} + \nu$, where $\gamma = 1.2$ and the probability of connection becomes roughly constant for distances greater than $\nu = 1000$km.

($ii$) Each agent can be in one of two states, susceptible ($S$) or infected ($I$).  Initial adoption is seeded to a small fraction of agents who are initialized as infected.  Spreading is modeled over a series of $T$ time periods, where the number of agents in each state is tracked (subject to $S(t) + I(t) = N$).  Each time period, all infected agents attempt to infect their neighbors.  With probabilities $\beta_{r}$ and $\beta_{e}$, a regular or early adopter, respectively, will heed a recommendation and adopt the technology.  We use the ratio, $R = \frac{\beta_e}{\beta_r}$ to control differences in early versus regular adopters.  These features mimic social dynamics that suggest the pressure to adopt increases as more friends adopt and that more connected people receive greater benefits from adopting social technologies\cite{Valente1995}.  Some models assume that an individual will adopt an innovation once a specific number \cite{Granovetter1973, Watts2002} or proportion \cite{Centola2007} of their contacts have also adopted. Others have found evidence that occupying similar roles in social networks is more predictive of adoption \cite{Burt1987}.  While we do not attempt to test these hypotheses, Kleinberg has suggested that the dynamics of these adoption schemes are quantitatively similar \cite{Kleinberg2007}.

($iii$) In addition to word-of-mouth spreading, we also incorporate a media agent.  This agent can be thought of either as an external force responsible for adoptions, similar to the Bass model, or as specialized agent that is connected to every other node in the social network.  Each time period, the media broadcasts its message to adopt a technology, and each agent flips a coin determining if adoption occurs.  The media transmission probability is given by, Pr$(media \ infection) = \alpha M$, where $\alpha$ is a model parameter, and $M$ is the endogenous media volume.  Media volume itself is determined as a function of the number of previously infected agents, $I(t-1)$, and a random term $\epsilon$ such that $M(t) = I(t-1) + \epsilon$.  For convenience, we normalize the media so that, $M(t) \in [0,1]$.  Finally, we set the size of random shocks $\epsilon$ to be on the order of $M(t)$, reflecting stylized features seen in Google News volume data.

\section{Results and Discussion}
\subsection{Replicating standard SI model}
We first present results for parameter settings that reduce our simulation to the traditional SI model.  We set $\beta_r = \beta_e$ (leaving only one type of agent), $\alpha = 0$ (removing the media), and populate each of $L=408$ cities uniformly with $1000$ agents for a total population of $N = 408,000$.  We initialize the network to have a completely random spatial distribution of links (i.e. $p_r = \frac{1}{r_{max}}.$) and a Poisson degree distribution. We choose a Poisson degree distribution because the structure of the adoption network is more selective than a scale free structure found in measurements of all connections in online social networks~\cite{Leskovec2007,Wu2004,Goncalves2011}. For example, Leskovec et al. \cite{Leskovec2007} found that individuals who recommended a product to tens or even hundreds of contacts influenced no more purchases on average than those who sent recommendations to just a few friends. 

Thus, we expect the number of people who can influence a person to adopt a technology is smaller than the number of acquaintances they have and the distribution is not likely to be long tailed.  Scaling these numbers to fit our simulation size we choose a reasonable average degree of $\langle k \rangle = 7$.

\begin{figure} 
\centering
	\includegraphics[width=1\linewidth]{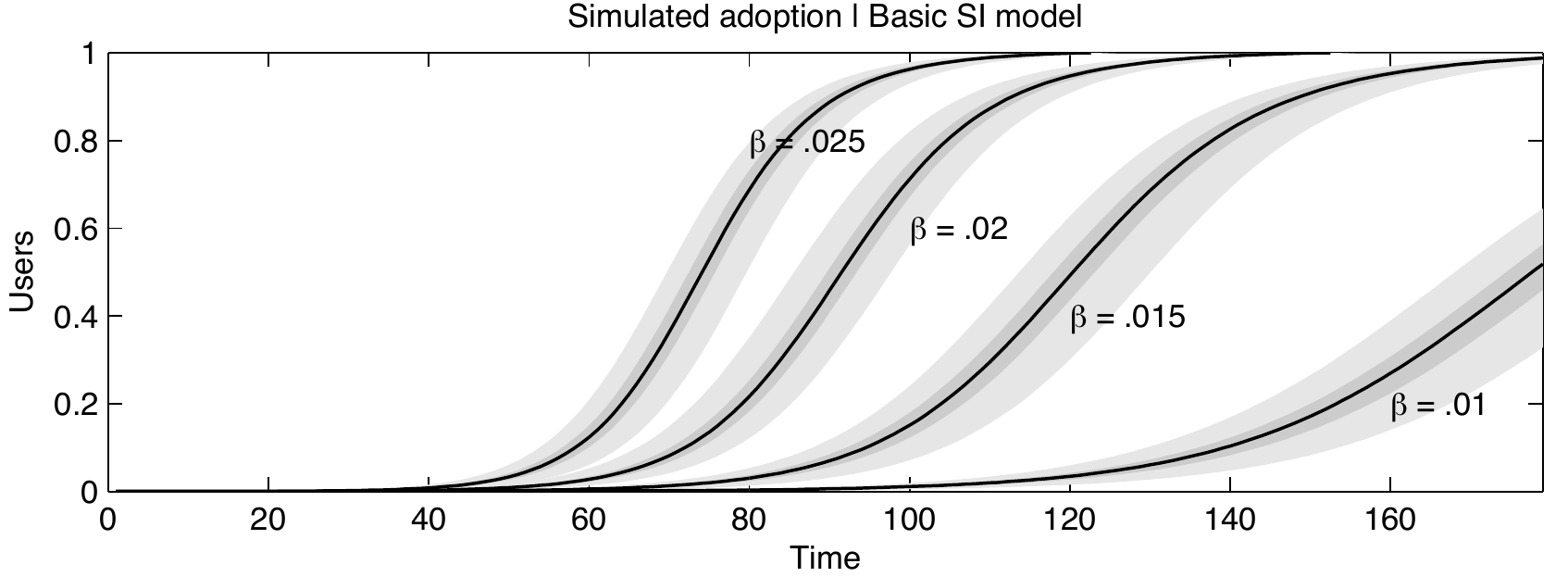} 
	\caption{To verify that our simulation replicates traditional results, we set parameters that mirror a classic SI model.  Four different transmission rates $\beta$ are displayed, each run 500 times and averaged.  The bands surrounding the average value are bounds containing 75\%, and 95\% of simulation runs.}\label{fig:sim_basicSI}
\end{figure}

Fig. \ref{fig:sim_basicSI} displays the simulated number of adopters per week for a variety of values for $\beta$.  The simulation was run 500 times for each of the parameter settings.  The bands surrounding the average represent ranges between which 75\% and 95\% of simulations fell. In this simple form of the model, it is not possible to reproduce the empirical shape of the cumulative adoption curve seen in the Twitter case study.

Next, we add more diverse geography to the model in the form of city populations, geographically distributed friendships, and early adopters that are three times as likely to adopt when than regular adopters ($R = \frac{\beta_e}{\beta_r} = 3$).  To understand how these additions affect adoption at the local level, we first examine the importance of network structure in the presence of two agent types.  

An important discovery regarding the social network of adoption is that the addition of the correct proportion of highly susceptible early adopters to cities is not enough to reproduce the observed trends.  A very specific type and strength of homopholy must be present to ensure that the early adopters are connected to each other, forming a giant component and not leaving members of their type isolated by regular adopters.  To form such a giant component, find that not only must agents prefer friendships with other agents of similar type ({\it geographically unbiased}, they must also prefer friendships with those closer to them geographically {\it geographically biased}. \footnotetext{Geographically biased friendships are selected as a function of distance with probability $p_r$ described in previous sections~\cite{Nowell2005}.} Together, these two sources of homopholy are enough to ensure a giant component of early adopters forms in the network.

\begin{figure} 
\centering
	\includegraphics[width=1\linewidth]{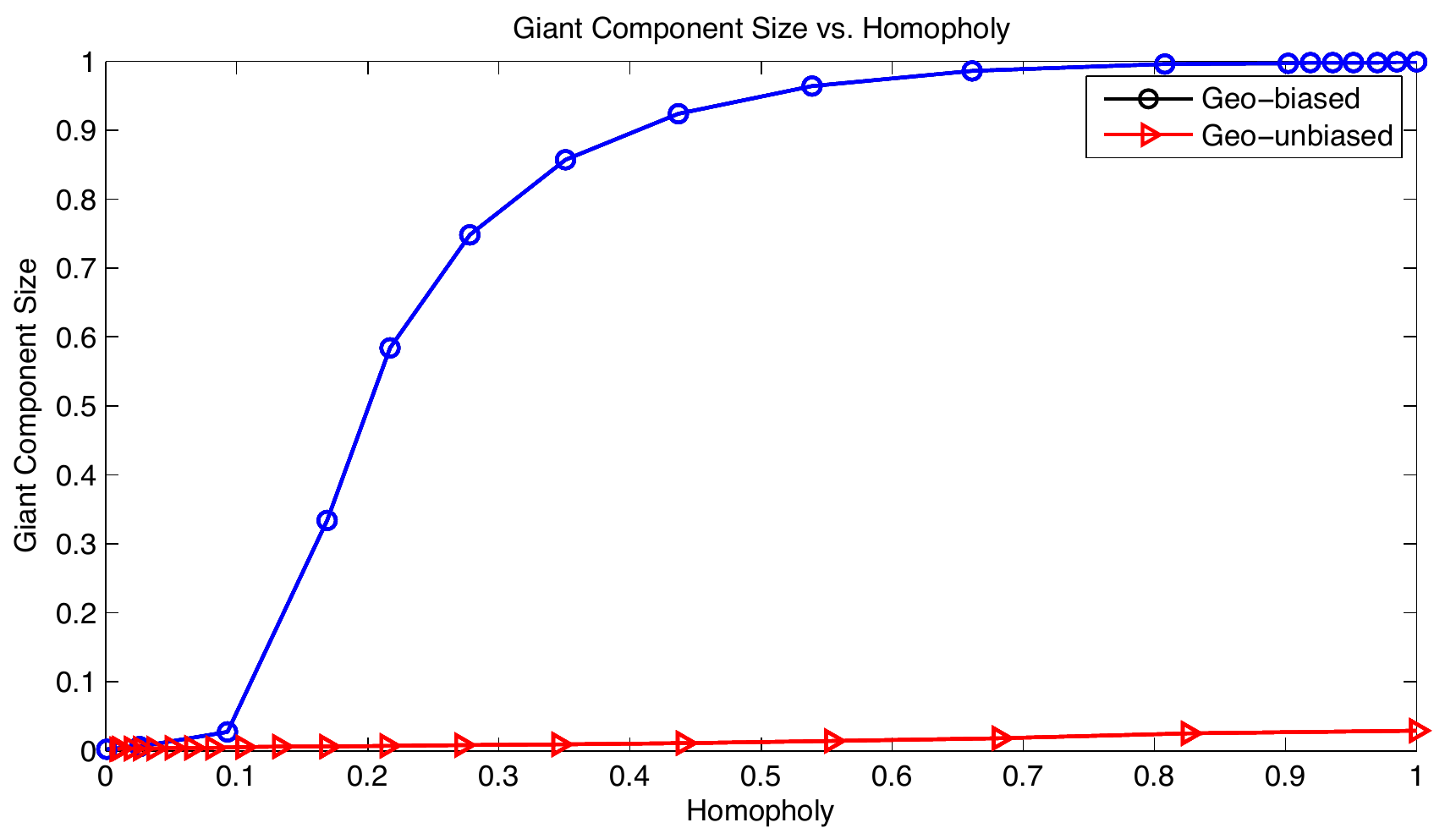} 
	\caption{The size of the giant component plotted against homopholy within the group of early adopters when friendships are biased or unbiased with respect to geography.  The figure illustrates preference for friendship with similar agents is not enough to connect early adopters and that geographically biased friendships are required.}\label{fig:comp_homoph}
\end{figure}

Fig. \ref{fig:comp_homoph} plots the size of the giant component of early adopters produced at a given level of homopholy among early adopting types for networks with either geographically biased or unbiased friendships.  Here we define homopholy as the average fraction of an early adopter's friends who are also early adopters.  These estimates were obtained by creating and consolidating results over $100$ networks, each with $N=10,000$ nodes and a given level of homopholy, then computing the size of the giant component. For the remainder of this paper, we refer to the scenario where there is enough homopholy, such that there exists a giant component containing over 95\% of the geographically biased early adopters.

\begin{figure*}
\includegraphics[width=1\linewidth]{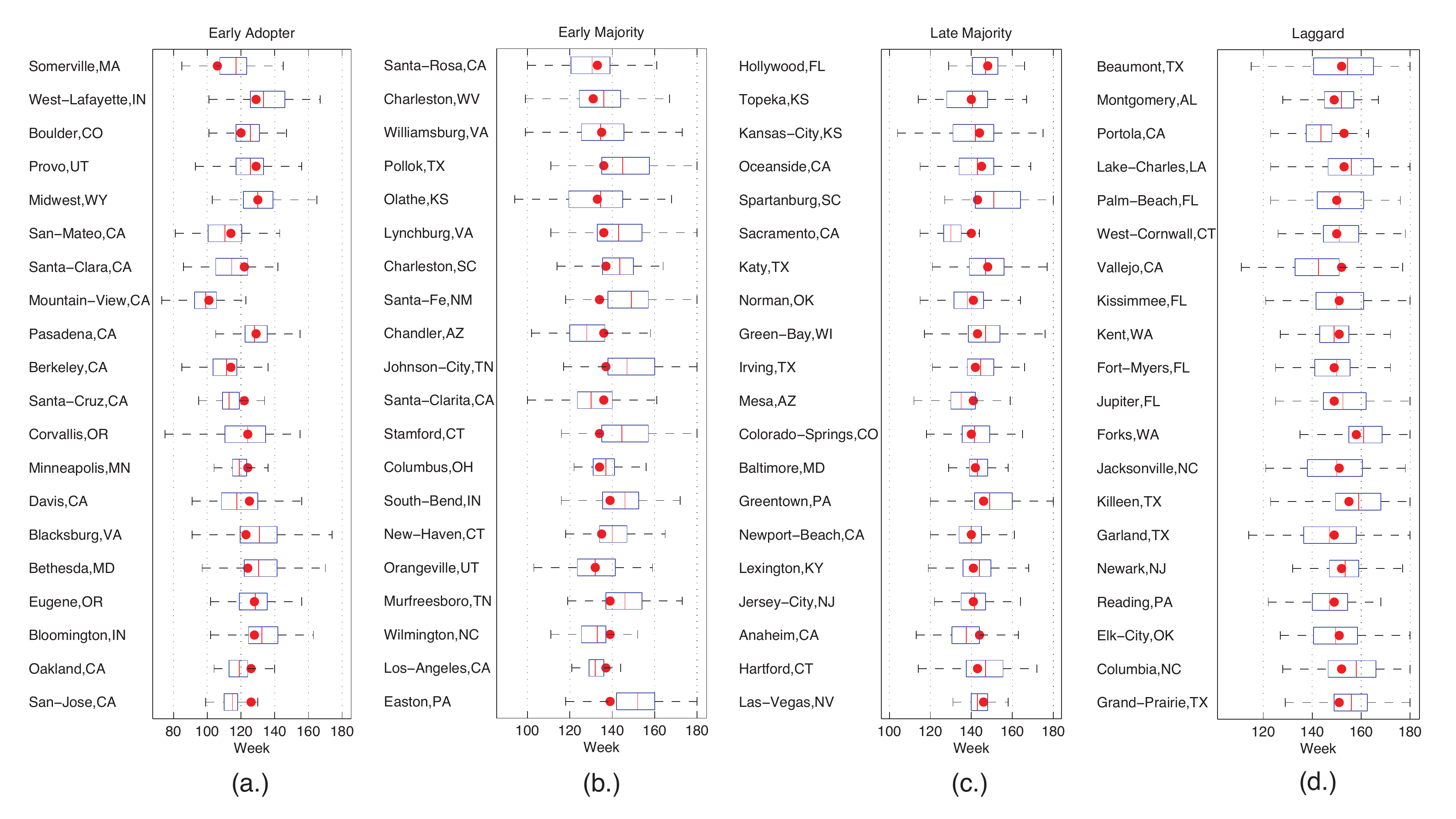} 
\caption{Simulation results are compared to actual critical mass achievement times for different subsets of locations.  Borrowing from the diffusion of innovations literature, we use four groups (a.) Early adopting, (b.) Early Majority,  (c.) Late Majority, (d.) Laggards.} \label{fig:boxplots}
\end{figure*}

\begin{figure} 
\centering
	\includegraphics[width=1\linewidth]{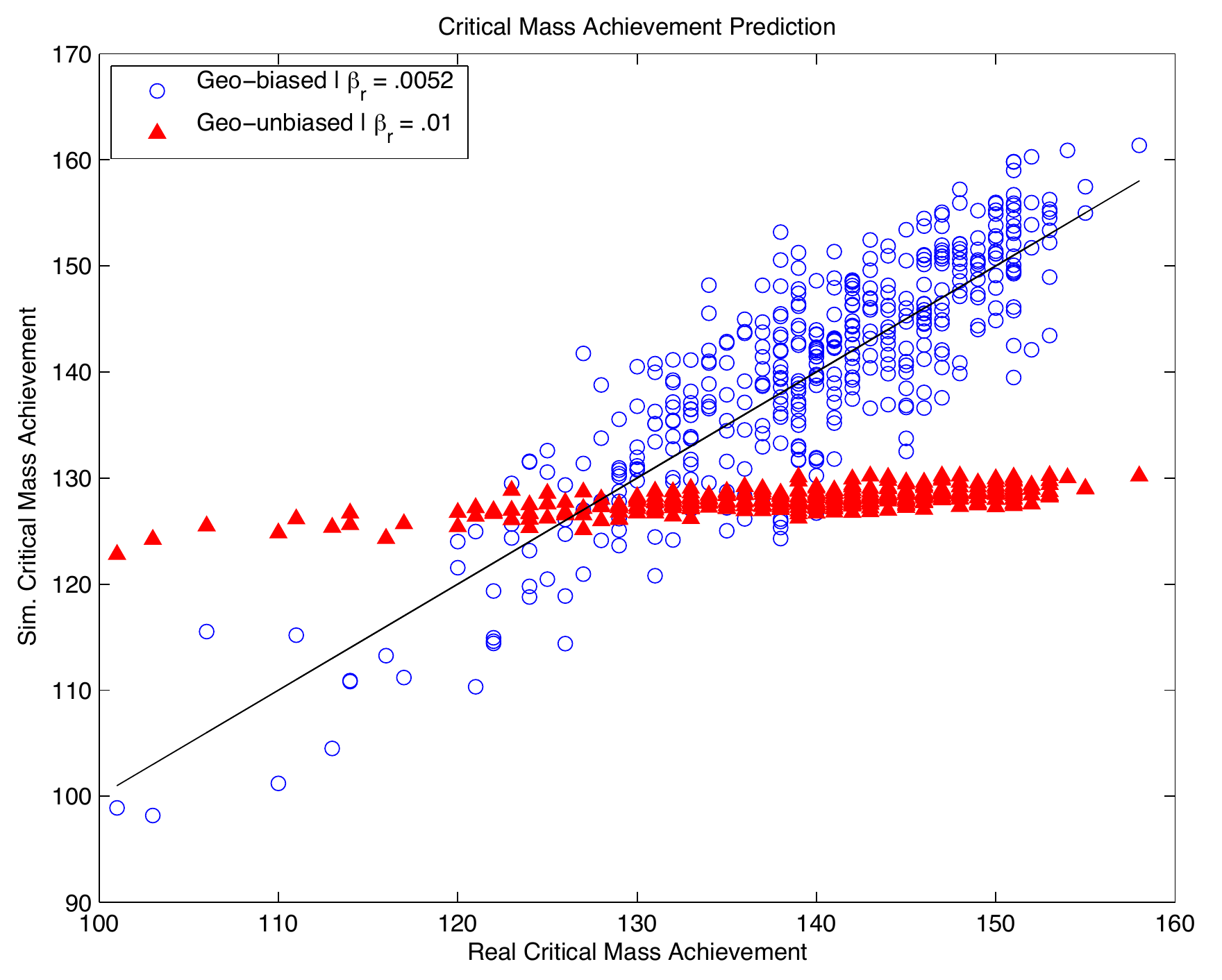} 
	\caption{Simulated critical mass achievement times are compared to times measured from Twitter data.  We find geographically biased friendships are critical to reproducing the intercity spread of Twitter.}\label{fig:CMAcomp}
\end{figure}

Fig.\ref{fig:CMAcomp} compares the predicted and actual times of critical mass achievement both with and without geographically biased friendships.  In the absence of a giant component, nearly all cities peak at the same time.  When geographically biased friendships are introduced such that a giant component of early adopters is formed, we are able to predict city level Twitter adoption, while preserving national trends. It is important to stress here that global cumulative adoption can be reproduced without the geographically biased network, and in most studies adoption is not geographically resolved. It is interesting to observe that just with the introduction of the spatial component in the network, it is possible to accurately simulate the critical mass achievement times in most cities. Fig. \ref{fig:boxplots} shows these simulated times when compared to times empirically measured in data.  We have divided specific cities into four groups based on when they reached critical mass relative to all locations.  For selected cities, simulation quartiles are plotted along with actual peak times\footnote{In the on-line supplementary information we provide data files containing the composition and adoption times of different cities, with the goal of facilitating future studies of other hypothesis and types of adoption.}.

\subsection{Media Influence}
\begin{figure} 
\centering
	\includegraphics[width=1\linewidth]{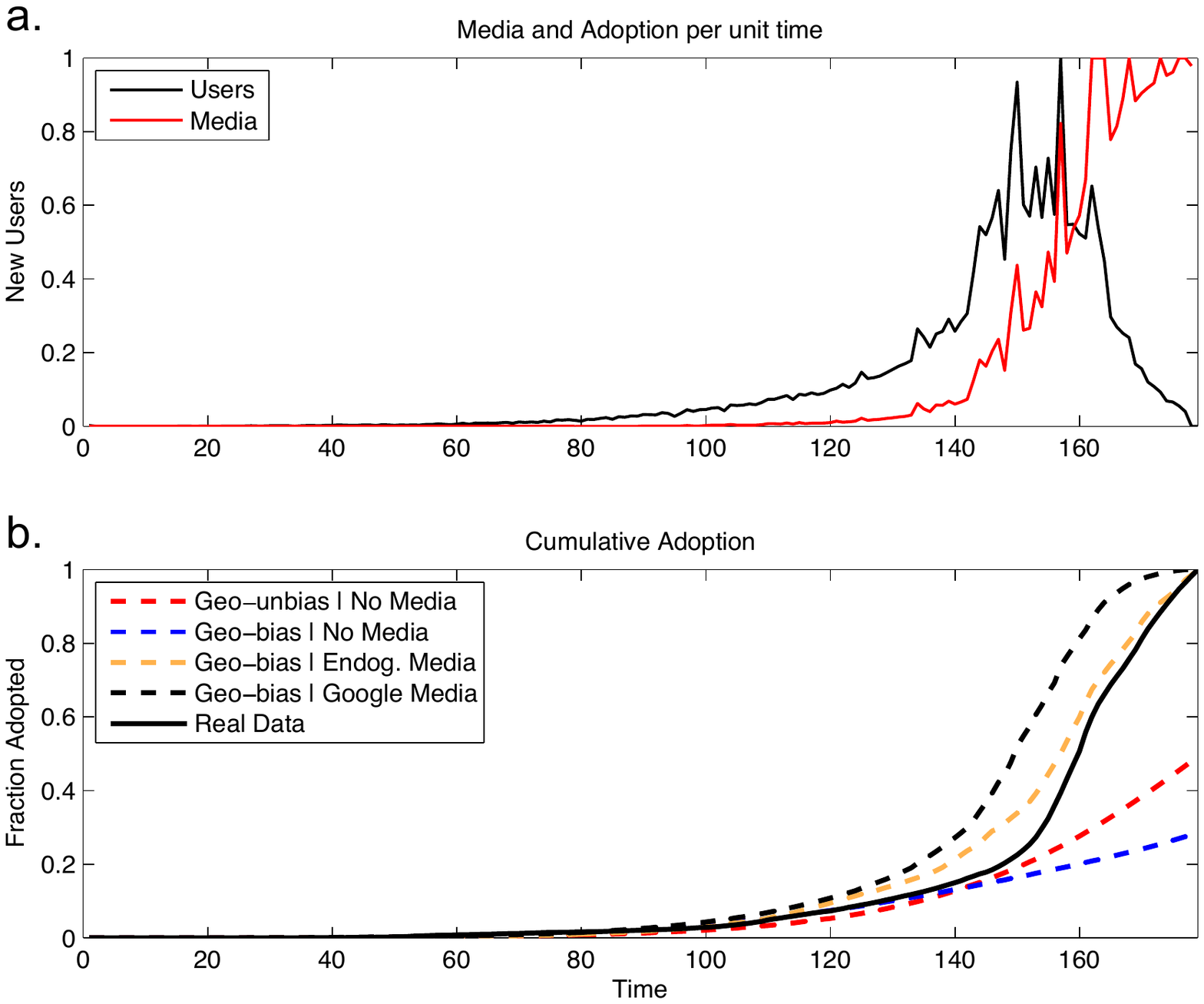} 
	\caption{Simulated adoption treating the media as endogenous and increasing with the number of adopters.  (a.) Shows simulated new users per week (normalized to the maximum over the period) as well as normalized media volume each week.  (b.) A comparison of all model scenarios.  Traditional models, excluding geographic bias as well as media perform the worst.  Including geographically biased friendships increases and media affects produce adoption predictions which more closely resemble real data.}\label{fig:endogmedia}
\end{figure}

Fig. \ref{fig:endogmedia} compares predictions of national adoption with the above model conditions.  Both the classic logistic growth curve predicted by geographically unbiased scenario, as well as the geographic biased case in friendships agree closely with the actual adoption of Twitter in the early stages of adoption.  For later stages, however, our predictions fall far short of the actual adoption (see Fig.~\ref{fig:endogmedia}).  Examining news volume as collected by Google, we notice that purely word-of-mouth simulations start performing badly around week 120 after launch, exactly when mass media begins to report on the web application. This transition allows us to measure the relative strength of word of mouth spreading versus mass media influence.

Predicting when the real media events occurred is beyond the scope of this work. We hence simulated adoptions introducing the empirical news volume from Google's database. In order to achieve the national adoption pattern similar to that seen in real data, we find that agents must be highly susceptible to media influence, with the parameter $\alpha = .15$.  Contrasting aggregate adoption predictions both with and without media influence suggests that the mass media was responsible for at least half of the newly joined Twitter's users, especially in later stages.  Coupled with our early results showing the importance of homopholy and geography during the early stages of spread, our model paints a much more complete picture of adoption, capable of reproducing both aggregate and local trends in space and time.

In general, as described above, this relatively simple model can treat news volume as endogenous such that adoption may be simulated without requiring external empirical data on media influence. Fig. \ref{fig:AggWeekTS} reveals three salient features of mass media volume largely ignored by traditional models.  The first feature suggests media coverage of a new product or trend is increasing in the number of users that have adopted and that this relationship is super-linear.  The second feature is that media volume does not accelerate until a critical mass of users has been reached.  Finally, news data suggests that random events such as celebrity endorsements or unpredictable political uprisings can have huge affects on media coverage, increasing news volume as much as two fold.

When implementing this model in our simulation, we include both agent types (early and regular adopters) and geographically biased contacts.  Fig. \ref{fig:endogmedia} displays an example of a simulation run with endogenous media when parameters were tuned to replicate features of Twitter's adoption.  Qualitatively, the simulated adoption curve is characterized by a relatively modest growth rate during early stages of adoption when the media is non-existent because there is no reason to report on a product.  Critical mass is reached such that local geographic spread is reproduced.  Shortly after critical mass is reached, media coverage begins increasing rapidly, influencing the rest of the agents to adopt.

\section{Conclusion}
In light of the globalized world accessible via the Internet, previous models of adoption fail to characterize the interplay of media and word of mouth.  In this article, we have presented descriptive statistics of the spatiotemporal adoption of a web application and proposed a model of technology adoption or, more generally, social contagion, to replicate features seen in data from city to national scales.  For early stages, when spreading occurs primarily through word-of-mouth, we find that adoption is strongly correlated with traditional demographic covariates.  Early adopting cities tend to be those with large, tech-savvy and younger populations such as Silicon Valley and universities.  Media influences during later stages, however, were found to be very strong, accounting for a two to four fold increase in the number of people who adopted.

Our model extends previous work in two important ways. First, we demonstrate that geographical bias of the social network is a crucial ingredient to reproduce the dynamic of adoption at a city scale.  The media features of the model captures our empirical observations that the media reacts to the number of adopters with a super-linear trend after a product has reached a critical mass and with random shocks that emulate super-influential or major media events.    
 
These results suggest that our model is capable of replicating both micro (at the city level) and macro (at the national level) adoption phenomena and may provide substantial improvement over existing frameworks such as the SI or Bass models.  We hope it inspires future work in the area.  Specifically, it is interesting to compare and contrast the spatial diffusion of web apps such as Twitter, with more tangible products such as gadgets, medicine, or cars.  For example, it may be possible to use the composition of the cities as characterized by the adoption of Twitter to predict or even try to accelerate the adoption of other related kinds of technological innovations.  This work also represents advances in models of spreading in networks where the roll of demographics, i.e. node attributes, as well as geography is critical for future predictions.  These insights may be particularly useful in modeling opinion spreading such as in elections and collective action.

\begin{acknowledgments}
This work was funded in part by the National Science Foundation through a Graduate Research Fellowship to Jameson L. Toole, and the awards from NEC Corporation Fund and the Solomon Buchsbaum Research Fund.
\end{acknowledgments}

\end{document}